%Paper: hep-th/9209048
%From: dafr@geom.umn.edu
%Date: Mon, 14 Sep 92 12:54:03 CDT

%%%%%%%%%%%%%%%%%%%%%%%%%%%%%%%%%%%%%%%%%%%%%%
% This paper requires amstex version 2.1.  If you do not have that, you can
% obtain it from the American Mathematical Society by telnetting to
% e-math.ams.com (IP number 130.44.1.100) and using the login and password
% ``e-math''. Instructions for ftp access can be obtained by following the
% instructions on the screen.  You will also need the amstex fonts.  If this
% fails, request a copy of the paper by sending email to Dan Freed
% (dafr@math.utexas.edu).
%
% There is an encapsulated postscript file included for Figure 1 at the end
% of this document.
%
\input amstex
\input amsppt.sty
%%%%%%%%%%%%%%%%%%%%%%%%%%%%%%%%%%%%%%%%%%%%%%%
\CenteredTagsOnSplits
\NoBlackBoxes
\def\today{\ifcase\month\or

January\or February\or March\or April\or May\or June\or
 July\or August\or September\or October\or November\or December\fi
 \space\number\day, \number\year}
\define\({\left(}
\define\){\right)}

\define\CC{{\Bbb C}}

\define\RR{{\Bbb R}}

\define\Tr{\operatorname{Tr}}
\define\ZZ{{\Bbb Z}}
\define\[{\left[}
\define\]{\right]}

\define\chiup{\raise.5ex\hbox{$\chi$}}
\define\cir{S^1}

\define\exertag #1#2{\removelastskip\bigskip\medskip\eightpoint\noindent%
\hbox{\rm\ignorespaces#2\unskip} #1.\ }

\define\inv{^{-1}}
\define\mstrut{^{\vphantom{1*\prime y}}}
\define\protag#1 #2{#2\ #1}

\define\res#1{\negmedspace\bigm|_{#1}}
\define\temsquare{\raise3.5pt\hbox{\boxed{ }}}

\define\theprotag#1 #2{#2~#1}

\define\zmod#1{\ZZ/#1\ZZ}

%%%%%%%%%%%%%%%%%%%%%%%%%%%%%%%%%%%%%%%%%%%%%%%%%%%%%%%%%%%%%%%%%%%%%%%%%%%%%%%
\def\bigstrut{\hbox{\vrule height18pt depth 9pt width0pt}}
\def\dashfill{\cleaders\hbox to .8em{\hss--\hss}\hfill}
\def\entry#1:#2:#3 {\bigstrut{#1}&{#2}&{#3}\cr}
\def\titlestrut{\hbox{\vrule height12pt depth 9pt width0pt}}
\define\Adgi{\Ad_{g\inv}}
\define\Ad{\operatorname{Ad}}
\define\Lk{L^{(k)}}
\define\Lt{L^{(3)}}
\define\TT{{\Bbb T}}
\define\ac#1#2{S_X(#1,#2)}
\define\acT{S_X(\Theta)}
\define\bW{\partial W}
\define\bX{\partial X}
\define\bY{\partial Y}
\define\conn#1{\Cal{A}_{#1}}
\define\form{\langle \cdot \rangle}
\define\gauge#1{{\Cal G}_{#1}}
\define\intLk{\int_{M/N}\Lk}
\define\sect#1{\Cal{S}_{#1}}
\define\tpi{2\pi i}
\define\zo{[0,1]}
%%%%%%%%%%%%%%%%%%%%%%%%%%%%%%%%%%%%%%%%%%%%%%%%%%%%%%%%%%%%%%%%%%%%%%%%%%%%%%
\NoRunningHeads
\refstyle{A}
\widestnumber\key{RSW}
	\topmatter
 \title\nofrills Locality and Integration in Topological Field Theory \endtitle
 \author Daniel S. Freed  \endauthor
 \thanks The author is supported by NSF grant DMS-8805684, an Alfred P.
Sloan Research Fellowship, a Presidential Young Investigators award,
and by the O'Donnell Foundation.\endthanks
 \thanks This is an expanded version of a talk given at the XIX International
Colloquium on Group Theoretical Methods in Physics, Salamanca, July, 1992.
It will be published by Ciemat in the series Anales de f\'\i sica,
monograf\'\i as.  \endthanks
 \affil Department of Mathematics \\ University of Texas at Austin\endaffil
 \address Department of Mathematics, University of Texas, Austin, TX
78712\endaddress
%\curraddr \endcurraddr
 \email dafr\@math.utexas.edu \endemail
 \date September 14, 1992\enddate
%\dedicatory \enddedicatory
	 \abstract
 We discuss the geometry of topological terms in classical actions, which by
themselves form the actions of topological field theories.  We first treat
the Chern-Simons action directly.  Then we explain how the geometry is best
understood via integration of {\it smooth Deligne cocycles\/}.  We conclude
with some remarks about the corresponding quantum theories in 3~and
4~dimensions.
 	\endabstract
	\endtopmatter

\document

\comment
lasteqno @ 20
\endcomment

\subhead
Introduction
\endsubhead

A fundamental property of a field theory is its {\it locality\/}.  This
roughly means that computations can be made locally in spacetime.  Or, to
make a global computation one can compute locally and combine the results.
The easiest way to do this is to cut a spacetime along codimension one
submanifolds.  For example, in a 3~dimensional theory we cut along smooth
2~dimensional surfaces.  We could contemplate cutting these surfaces along
1~dimensional circles, which amounts to dividing up the original
3~dimensional manifold into manifolds with corners.  (A simple example would
be cutting a solid 3~dimensional ball along an equatorial disk.)  Such
computations have been done recently in {\it topological\/} field theories,
and they lead to a variety of interesting mathematical results.  We are
therefore led to study the notion of locality in field theory more closely.

By its very nature a field theory is local.  The variables are local fields
on a manifold.  The {\it lagrangian\/} is then a local function of these
fields and the classical {\it action\/} is the integral of these fields over
spacetime.  The quantum theory is also defined using integration, this time
over an infinite number of variables.  Such {\it path integrals\/} are not
well-understood by mathematicians.  Recent axiomatizations of the path
integral in conformal field theory~\cite{S} and in topological field
theory~\cite{A} express the fundamental nature of the integral in terms of
{\it gluing laws\/}.  They describe how the path integral behaves when we cut
along codimension one manifolds.  There are corresponding gluing laws for the
classical action, which after all is also an integral, but typically this
simply expresses a basic property of ordinary integration.

Topological actions, such as the Wess-Zumino-Witten term in 2~dimensions or
the Chern-Simons action in 3~dimensions, lead to more interesting geometry
and topology, even in the classical theory.  Such actions were studied by
Alvarez~\cite{Alv} and more recently by Gaw\c edzki~\cite{G}, who introduced
{\it Deligne cocycles\/} in this context.  We develop these ideas more fully
in~\cite{F2}, ~\cite{F3}, where we construct a complete integration theory
for what we call {\it higher line bundles with connection\/}.  (These objects
have appeared in other contexts recently in the work of Brylinski~\cite{Br}.)
Our purpose in this note is to explain that topological actions, and their
associated geometry, are best understood in the context of this integration
theory.  Our discussion centers around the 3~dimensional Chern-Simons theory.
We do not begin immediately with the higher line bundles, but rather first
develop some of the geometry in a more familiar way.  (See~\cite{F1} for more
details.)  This applies in the gauge theory to connections on trivializable
bundles.  Then we give a brief introduction to the integration theory of
higher line bundles and show how it applies to topological actions.

We also find a topological ``classical theory'' in 4~dimensions which is
related to the Chern-Simons functional (see Table~3).  The action is the
usual characteristic number for a bundle over a 4-manifold.  One can
speculate that this is an appropriate classical theory for the 4~dimensional
topological quantum field theory which computes the Donaldson polynomial, but
since the path integral here is so vague it is hard to put much stock in such
statements.  This classical action is in some sense constant over the space
of connections on a fixed bundle, so it is hard to see what to integrate
over!  Still, we find the analogies suggestive, so we describe them briefly
at the end of this note.

\subhead
The Chern-Simons lagrangian
\endsubhead

The basic data we need to define the Chern-Simons action is
  $$ \aligned
     G\qquad &\text{compact Lie group}\\
     \lambda \in H^4(BG)\qquad &\text{integral cohomology class} \endaligned
     \tag{1}$$
We restrict here to groups~$G$ which are {\it connected\/} and {\it simply
connected\/}.  As a result we won't need to worry in this lecture about the
classifying space~$BG$ or the class~$\lambda $.  All we need to know is that
$\lambda $~determines an invariant symmetric bilinear form
  $$ \form\:\frak{g}\otimes \frak{g}\longrightarrow \RR \tag{2}$$
on the Lie algebra~$\frak{g}$.  Also, this form is {\it integral\/} in a
certain sense.  The prime example is~$G=SU_2$ with bilinear form
  $$ \langle a\otimes b  \rangle = -\frac{k}{8\pi ^2}\Tr(ab),\qquad a,b\in
     \frak{s}\frak{u}_2, $$
where $k$~is some integer and $a,b$~are $2\times 2$~traceless skew-hermitian
matrices.  There is a canonical left-invariant $\frak{g}$-valued
1-form~$\theta $ on~$G$, the {\it Maurer-Cartan form\/}, usually
written~$\theta =g\inv dg$.  From~$\theta $ and~$\form$ we construct an
ordinary 3-form
  $$ -\frac16 \langle \theta \wedge [\theta\wedge
     \theta]\rangle \in \Omega ^3_G. \tag{3}$$
The integrality of~$\form$ implies that \thetag{3}~has integral periods;
that is, its integral over closed 3-cycles in~$G$ is an integer.

A gauge field on a manifold~$M$ is a connection on a principal $G$~bundle
over~$M$.  That is, $P$~is a manifold with a free (right) $G$~action, and the
space of orbits is diffeomorphic to~$M$.  A {\it connection\/}~$\Theta \in
\Omega ^1_P(\frak{g})$ is a $\frak{g}$-valued 1-form on~$P$ which satisfies
certain affine equations.  Its {\it curvature\/}
  $$ \Omega =\Omega (\Theta ) = d\Theta  + \frac 12 [\Theta \wedge \Theta ]
     $$
is a $\frak{g}$-valued 2-form on~$P$.  From $\Theta $, $\Omega $, and the
bilinear form~$\form$ we construct two (scalar) differential forms:
  $$ \alignat 2
     \langle \Omega \wedge \Omega   \rangle&\in \Omega ^4_M&\qquad
     &\text{Chern-Weil form}  \tag{4}\\
     \alpha  = \alpha (\Theta ) = \langle \Theta \wedge \Omega \rangle -
     \frac{1}{6}\langle \Theta \wedge [ \Theta \wedge \Theta ]\rangle &\in
     \Omega ^3_P &\qquad  &\text{Chern-Simons form}   \tag{5}\endalignat $$
As written the Chern-Weil form lives on~$P$, but it pulls back from a 4-form
on~$M$.  The Chern-Weil form is closed, whereas the differential of the
Chern-Simons form is
  $$ d\alpha = \langle \Omega \wedge \Omega   \rangle. \tag{6}$$

Connections are local fields in the sense that they can be cut and pasted (or
glued).  The Chern-Simons and Chern-Weil forms are local ``functions'' of the
connection in the sense that they commute with cutting and pasting.  For
example, the Chern-Simons form of a glued connection is the gluing together
of the Chern-Simons form of the original connection.  As such these forms are
valid {\it lagrangians\/} for a field theory.  In such field theories the
integral of the lagrangian over spacetime is the {\it action\/}.  We examine
these integrals now in various dimensions.

\subhead
4 dimensions: Characteristic numbers
\endsubhead

Since the Chern-Weil form~\thetag{4} is a form on the base manifold, we can
integrate it if the base manifold is a compact oriented 4-manifold~$W$.  For
such a manifold we define
  $$ \lambda _W(\Theta ) = \int_{W}\langle \Omega (\Theta )\wedge \Omega
     (\Theta )  \rangle. \tag{7}$$
If $W$~is {\it closed\/}, i.e., $\partial W=\emptyset $, then the integrality
of the bilinear form~\thetag{2} (which comes from an integral cohomology
class~\thetag{1}) implies that $\lambda _W(\Theta )$~is an integer.  Since
\thetag{7}~is clearly a continuous function of~$\Theta $, and since the
space~$\conn P$ of connections on~$P$ is connected, the function~$\lambda _W$
is constant on~$\conn P$.  Its value is a topological invariant of~$P$, a
{\it characteristic number\/} or {\it topological charge\/}.  If $\partial
W\not= \emptyset $ then $\lambda _W$~takes real values and varies smoothly
with the connection~$\Theta $.

The locality of~$\lambda _W$ is expressed by the following {\it gluing
law\/}:
  $$ \lambda _W(\Theta ) = \lambda _{W_1}(\Theta _1) + \lambda _{W_2}(\Theta
     _2). \tag{8}$$

\bigskip
\midinsert
% FIGURE 1 GOES HERE
\nobreak
\centerline{Figure~1: Gluing manifolds and connections}
\bigskip
\endinsert

 \flushpar
 Here $\Theta _i$ is a connection on~$W_i$ and $\Theta _1$~and $\Theta
_2$~agree on the common boundary, so paste together into a connection~$\Theta
$ on~$W$.  This is the first of the gluing laws we shall see.  Here it
expresses a standard property of integration of differential forms.

\subhead
3~dimensions: The Chern-Simons invariant
\endsubhead

Now suppose $P\to X$ is a principal $G$~bundle over a compact oriented
3-manifold~$X$.  Let $\Theta $~be a connection on~$P$.  We would like to
integrate the Chern-Simons form~\thetag{5} over~$X$, but unfortunately that
form lives on~$P$, not~$X$.  So we introduce an auxiliary piece of data,
namely a section $p\:X\to P$ of the bundle $P\to X$, to pull down the
Chern-Simons form to~$X$.  Such sections exist because of our topological
hypotheses on~$G$.\footnote{We assume that $G$~is connected and simply
connected.  For general groups we need a more complicated construction with
the classification space at this point.  To carry this through to construct
the smooth Chern-Simons action we require the higher line bundles and Deligne
cocycles which we introduce later.} Hence define
  $$ \ac p \Theta = \int_{X}p^*\alpha (\Theta ). \tag{9}$$
Of course, $\ac p \Theta $~is a real number.  To get rid of the auxiliary
field~$p$ we investigate the dependence of~\thetag{9} on~$p$.  Now any
section of~$P$ can be written~$p\cdot g$ for some $g\:X\to G$, which can be
thought of as a gauge transformation.  Then we compute
  $$ \ac {p\cdot g}\Theta = \ac p \Theta
     + \int_{\partial X}\langle \Adgi p^*\Theta \wedge g^*\theta \rangle +
     \int_{X}g^*\!\(-\frac16 \langle \theta \wedge [\theta\wedge
     \theta]\rangle \). \tag{10}$$
Recall that $\theta =g\inv dg$ is the Maurer-Cartan 1-form on~$G$.

Suppose first that $X$~is closed, i.e., $\bX=\emptyset $.  Then the second
term in~\thetag{10} vanishes.  Also, the third term is an integer by the
integrality of the form~\thetag{3}.  So up to an integer the
action~\thetag{9} is independent of~$p$.  (This is a well-known argument.)
It is natural, then, to exponentiate and define
  $$ e^{ 2\pi i\acT} = e^{ 2\pi i\ac \Theta p}\quad \text{for any section
     $p$}. \tag{11}$$
This action takes values in the circle group
  $$ \TT = \{\lambda \in \CC : |\lambda |=1\}. $$

If $\bX\not= \emptyset $ then \thetag{10}~shows that $\ac\Theta p$~does
depend on~$p$.  However, if we reduce modulo integers again, then we see that
the dependence is only through the boundary values of~$p$, $g$, and~$\Theta
$.  That assertion is clear for the second term (even without reducing modulo
integers), and follows for the third term from the integrality
of~\thetag{3}.  This term is usually called the {\it Wess-Zumino-Witten
term\/}, and we denote it as
  $$ W_{\bX}(\partial g) =  \int_{X} g^*\!\(-\frac16 \langle \theta \wedge
     [\theta\wedge \theta]\rangle \) \pmod1 . $$
Let's rewrite these terms directly in terms of the boundary.  In fact, they
make sense for any connection~$\eta $ on a $G$~bundle $Q\to Y$ over any
closed oriented 2-manifold~$Y$.  Let $q\:Y\to Q$ be a section and set
$a=q^*\eta $; then $a\in \Omega ^1_Y(\frak{g})$ is a $\frak{g}$-valued 1-form
on~$Y$.  For any map $h\:Y\to G$ we set
  $$ c\mstrut _Y(a,h) = \exp\( \tpi\int_{Y}\langle \Ad_{h\inv } a\wedge
     h^*\theta \rangle + W_Y(h)\). \tag{12}$$
We rewrite~\thetag{10} in the form
  $$ e^{\tpi\ac\Theta {p\cdot g}} = e^{\tpi\ac\Theta p}\cdot
     c\mstrut _{\bX}\((\partial p)^*\partial \Theta ,\partial g\), \tag{13}$$
where $\partial p$, $\partial g$, and $\partial \Theta $ are the boundary
values of $p$, $g$, and~$\Theta $.  In other words, as a function of~$p$ the
exponentiated action is exactly determined by the ``cocycle''~\thetag{12}
once we know its value for a single section.

In typical mathematical fashion we introduce the space of all such functions
of~$p$.  Since the dependence is only through the boundary values this
construction applies to any connection~$\eta $ on $Q\to Y$ over a closed
oriented 2-manifold~$Y$ as above.  Let $\sect Q$ be the space of sections
$q\:Y\to Q$.  Then set
  $$ L_\eta  = L_{Y,\eta } = \{f\:\sect Q\to\TT : f(q\cdot h) = f(q)\cdot
     c_{Y}(q^*\eta ,h) \quad \text{for all $h\:Y\to G$} \}. \tag{14}$$
Again, a function~$f$ in~$L_\eta $ is determined by its value at any
section~$q_0\in \sect Q$, and specifying~$q_0$ gives a trivialization $L_\eta
\cong \TT$.  But without specifying a section~$q_0$ we cannot
identify~$L_\eta $ with~$\TT$.  Rather, it is a principal homogeneous space
for~$\TT$: Any $\lambda \in \TT$ multiplies any function~$f\in L_\eta $
yielding a new function~$f\cdot \lambda \in L_\eta $, and any two functions
in~$L_\eta $ are related by a unique such~$\lambda $.  Alternatively, we can
identify~$L_\eta $ as the elements of unit norm in a one dimensional complex
vector space with a hermitian metric; we simply view a function~$f\in L_\eta
$ as complex-valued.  We will fluidly pass between these interpretations
of~$L_\eta $ and between the names ``Chern-Simons circle'' and ``Chern-Simons
line'' for this space.  Equation~\thetag{13} now asserts
  $$ e^{\tpi \acT} \in L_{\partial \Theta }. \tag{15}$$
The construction of the Chern-Simons line appears in~~\cite{RSW}.

The Chern-Simons invariant~\thetag{11}, \thetag{15} is the action (or a term
in the action) of a field theory.  It differs from typical actions in a few
ways.  First, a typical action has values in the real numbers both on closed
spacetimes and on spacetimes with boundary.  The Chern-Simons action is only
well-defined modulo integers on closed spacetimes and takes values in a
principal $\TT$-space for spacetimes with boundary.  Also, a typical action
depends on a fixed background metric and is only invariant under
transformations of spacetime which preserve the metric, i.e., isometries.
The Chern-Simons action is defined without introducing a metric and is
invariant under all orientation-preserving diffeomorphisms.\footnote{It is
also invariant under gauge transformations, but these are symmetries of the
fields, not of the spacetimes, so play a different role.} The characteristic
property of any action is its locality, expressed through a gluing law as
in~\thetag{8}.  For the Chern-Simons action the gluing law takes the form
  $$ e^{\tpi\acT} = e^{\tpi S_{X_1}(\Theta _1)}\cdot e^{\tpi S_{X_2}(\Theta
     _2)}, \tag{16}$$
where $\Theta _1$, $\Theta _2$~are connections on~$X_1$, $X_2$ whose boundary
values agree along the common boundary, as pictured in Figure~1.  Since these
boundaries are identified with {\it opposite\/} orientations, the
Chern-Simons lines in which the Chern-Simons actions on the right hand side
of~\thetag{16} take their values are dual.  The~`$\cdot $' in~\thetag{16} is
the duality pairing of these Chern-Simons lines.

\subhead
2 dimensions: The Chern-Simons circle bundle
\endsubhead

In the previous section we constructed a principal $\TT$-space~$L_\eta $ for
every connection~$\eta $ over a closed oriented 2-manifold.  Now we
consider~$L_\eta $ as a function of~$\eta $.  Namely, fix a $G$ bundle~$Q\to
Y$ and let $\conn Q$ denote the affine space of connections on~$Q$.  Then the
circles~$L_\eta $ fit together into a {\it smooth\/} circle bundle
  $$ L_Q\longrightarrow \conn Q. \tag{17}$$
We call this the Chern-Simons circle bundle (or line bundle).  The smoothness
follows from the smooth dependence of~\thetag{12} on the first variable.
Also, the construction naturally leads to an action of the group of gauge
transformations~$\gauge Q$ on~$L_Q$.

Now suppose $\eta _t,\,0\le t\le 1$, is a path of connections on~$Q$.  This
path defines a connection with no $dt$~component on the bundle $\zo\times
Q\to\zo\times Y$.  Since $\partial (\zo\times Y) = -\{0\}\times Y \;\cup\;
\{1\}\times Y$ (the minus sign indicates the orientation), the Chern-Simons
action of this connection is naturally a map $L_{\eta _0}\to L_{\eta _1}$.
The diffeomorphism invariance of the Chern-Simons action shows that this map
is independent of the parametrization of the path.  Furthermore, if we glue
together two paths, then (a generalization of) the gluing law~\thetag{16}
implies that the corresponding maps on the Chern-Simons circles compose.  In
other words, the Chern-Simons action defines the parallel transport for a
connection on the circle bundle~\thetag{17}.

This geometric structure---a connection on the Chern-Simons circle
bundle~\thetag{17}---is an important part of the classical theory.  For
example, its curvature
  $$ \omega (\dot{\eta}_1 ,\dot{\eta}_2 )= - 2\int_{Y}\langle \dot{\eta}_1
     \wedge \dot{\eta}_2 \rangle ,\qquad \dot{\eta}_1 ,\dot{\eta}_2 \in
     T_\eta \conn Q, $$
is a symplectic form on~$\conn Q$.  This derivation of the symplectic form is
a geometric rendition of the standard link between Lagrangian mechanics and
Hamiltonian mechanics.  Notice that the hamiltonian function in this theory
vanishes; there is no local motion.  This is a hallmark of topological
theories.

The lift of the $\gauge Q$~action to~$L_Q$ preserves the connection on~$L_Q$
and so defines a {\it moment map\/} for the $\gauge Q$~action on the
symplectic manifold~$\conn Q$.  From this geometric point of view the moment
map is the obstruction to descending the connection on~$L_Q$ to~$\conn
Q/\gauge Q$; it only descends over the zeros of the moment map.  In this
case the moment map is the curvature, so the Chern-Simons circle
bundle~\thetag{17} together with its connection descend to a circle bundle
with connection
  $$ L_Q\longrightarrow \Cal{M}_Q $$
over the moduli space of gauge equivalence classes of {\it flat\/}
connections.  This bundle with connection enters in the construction of the
quantum theory~\cite{W1}.\footnote{Notice that it is this Chern-Simons line
bundle, not the determinant line bundle, which is relevant here.}

\bigskip
\midinsert
{\setbox0\vbox{\kern6pt\hsize 200pt  \centerline{A section~$e^{\tpi S_X(\cdot
)}$ of the bundle~$r^*L_{\bX}$:\hfil} \abovedisplayskip=6pt
\belowdisplayskip=6pt $$ \CD
r^*L_{\bX}  @>>> L_{\bX}\\ @VVV @VVV \\ \conn X @>r>> \conn{\bX}\endCD$$}
\offinterlineskip \tabskip = 0pt
  \halign{
	\vrule\enspace\hfil#\hfil\enspace\vrule\hskip2pt
        \vrule&\enspace\hfil#\hfil\enspace
	&\vrule\enspace\hfil#\hfil\enspace\vrule\cr
\noalign{\hrule}
\titlestrut{\bf dim}&{\bf closed manifold}&{\bf compact manifold with
boundary} \cr
\noalign{\hrule}
\vphantom{\vrule height 2pt}&&\cr \noalign{\hrule}
\entry 4:{A function $\lambda_W:\conn W\to\ZZ$}:{A function $\lambda_W:\conn
W\to\RR$}
\omit\vrule\dashfill\vrule\hskip2pt
\vrule&\omit\dashfill&\omit\vrule\dashfill\vrule\cr
\entry 3:{A function $e^{2\pi i S_X(\cdot)}\:\conn
X\to\TT$}:{$\vcenter{\box0}$} \noalign{\hrule}
\entry 2:{$\vcenter{\vbox{\kern6pt\hsize 170pt \centerline{A circle
bundle with connection}\vskip12pt\centerline{$L_Y\to\conn Y$}\kern6pt}}$}:{}
\vphantom{\vrule height2pt}&&\cr \noalign{\hrule}}
}
\nobreak
\bigskip
 \centerline{Table~1: Constructions with the Chern-Simons lagrangian
\footnotemark}
 \medskip
\endinsert

We summarize our constructions so far in Table~1.  We remark that the first
line in Table~1 fits better into a different table (Table~3), as we explain
later.  Looking at that table it is natural to ask whether one can construct
a Chern-Simons circle for connections on compact oriented 2-manifolds {\it
with boundary\/}, and whether there is a gluing law for such circles.  In
other words, is the Chern-Simons circle (and more generally the Chern-Simons
circle bundle~\thetag{17} with its connection) {\it local\/} on the surface?
In~\cite{F1,\S4} we carry out such a construction for fixed values of the
boundary holonomy.  Without fixing any data on the boundary, however, we are
forced to look at more abstract objects than principal $\TT$-spaces or
complex lines.  That motivates our introduction of ``higher line bundles''
\footnotetext{Our notation in this and subsequent tables is somewhat
lax. Instead of~`$\conn X$' we should use~`$\conn P$' for connections on a
fixed $G$~bundle $P\to X$, or perhaps~`$\overline{\conn X}$' for equivalence
classes of connections (connections modulo gauge equivalence) on~$X$.}
and ``Deligne cocycles''.\footnote{These higher line bundles and Deligne
cocycles already enter the construction of the Chern-Simons circle bundle
$L_Q\to\conn Q$ associated to {\it nontrivial\/} bundles $Q\to Y$ over closed
oriented 2-manifolds.  These nontrivial $G$~bundles exist if $G$~is not
connected or is not simply connected.  I do not know a construction of these
Chern-Simons circle bundles and of the action on 3-manifolds with boundary
which avoids this.}

\subhead
Higher line bundles, Deligne cocycles, and integration
\endsubhead

Our main thesis in this talk is that the Chern-Simons action, and other
topological actions like the Wess-Zumino-Witten action, are best understood
as integration of a differential-geometric object which we term a ``higher
line bundle with connection''.  (More accurately we should say instead
``higher circle bundle with connection'' or ``higher hermitian line bundle
with connection''.  We fluidly pass back and forth between the words ``higher
line bundle'' and ``higher circle bundle'', though this is slightly
misleading.)  Such objects are represented in \v Cech theory by ``Deligne
cocycles'', just as a circle bundle with connection is represented in \v Cech
theory by transition functions~$g_{ij}$ and local connection 1-forms~$\alpha
_i$.  One should think of a higher circle bundle with connection as a local
differential-geometric object, much like a differential form.  For example,
these animals pull back under maps and glue together in the way differential
forms do.  The new ingredient in all of this is an integration theory for
these higher circle bundles with connection.  It has all of the properties
one expects by analogy with integration of differential forms: diffeomorphism
invariance, sensitivity to orientation, Stokes' theorems, and a gluing law.
The gluing law expresses the locality of the integration process.  However,
there are two crucial conceptual differences between higher circle bundles
with connection and differential forms.  First, a differential form is a
section of a vector bundle, whereas a higher circle bundle is not.  Nor is a
higher circle bundle a fiber bundle over the base; there is no total space
which makes sense as a manifold.  Second, a differential form does not have
automorphisms whereas a higher circle bundle does.  It is crucial to keep
track of these automorphisms when gluing or considering group actions.  A
closely related statement is that it does {\it not\/} suffice for us to
consider equivalence classes of higher circle bundles with connection.  (These
equivalence classes are the {\it smooth Deligne cohomology\/}.)  There is
another reason it does not suffice to consider equivalence (or cohomology)
classes---one cannot integrate cohomology classes over manifolds with
boundary.  This is also true with differential forms---one cannot integrate
de Rham cohomology classes over manifolds with boundary.

After all of this hype it is fair to ask: What is a higher circle bundle with
connection?  We begin to answer that question by tabulating the lowest degree
cases in Table~2.  A few more hints: The \v Cech representative of an element
in the second column is a Deligne cochain which is not necessarily closed.
The objects in both columns have a ``curvature'' which is a $(k+1)$-form
on~$M$.  The curvature is closed for objects in the left hand column.  The
objects in the left hand column have a characteristic class in the integral
$(k+1)$-cohomology of~$M$.

\bigskip
\midinsert
{\offinterlineskip \tabskip = 0pt
\setbox0\vbox{\kern6pt\hsize 200pt\centerline{A section of a circle bundle
with connection}\vskip12pt\centerline{$L\longrightarrow M$}\kern6pt}
\halign{
	\vrule\enspace\hfil#\hfil\enspace\vrule\hskip2pt
        \vrule&\enspace\hfil#\hfil\enspace
	&\vrule\enspace\hfil#\hfil\enspace\vrule\cr
\noalign{\hrule}
\titlestrut{\bf degree~$k$}&{$\vcenter{\vbox{\kern6pt\hsize150pt
\bf \centerline{$k$-circle bundle with}\vskip6pt\centerline{connection
over~$M$}\kern6pt}}$}&{$\vcenter{\vbox{\kern6pt\hsize200pt
\bf \centerline{section of a $(k+1)$-circle bundle}\vskip6pt\centerline{with
connection over~$M$}\kern6pt}}$}
\cr
\noalign{\hrule}
\vphantom{\vrule height 2pt}&&\cr \noalign{\hrule}
\entry 0:{A function $M\to\TT$}:{$\vcenter{\box0}$}
\noalign{\hrule}
\entry 1:{$\vcenter{\vbox{\kern6pt\hsize 150pt \centerline{A circle
bundle with connection}\vskip12pt\centerline{$L\to M$}\kern6pt}}$}:{}
\vphantom{\vrule height2pt}&&\cr \noalign{\hrule}}
}
\nobreak
\bigskip
\centerline{Table~2: Low degree higher circle bundles}
\medskip
\endinsert

We still haven't said what these higher objects are!\footnote{See~\cite{F2}
for the formal definition.} From our point of view the best answer to that
question is in terms of Deligne cocycles---heuristically, a local
trivialization of a $k$-circle bundle with connection gives a Deligne
$k$-cocycle.  Let's first review the case~$k=1$ of an ordinary circle bundle
$L\to M$ with connection~$\alpha \in i\Omega ^1_L$.  Then an open
cover~$\{U_i\}$ of~$M$ and local trivializations $s_i\:U_i\to L\res{U_i}$
give rise to transition functions and local connection forms:
  $$ \gathered
     g\mstrut _{ij}=s_i\inv s\mstrut _j\: U_i\cap U_j\longrightarrow \TT\\
     \alpha _i=s_i^*\alpha \in i\Omega ^1(U_i). \endgathered \tag{18} $$
These satisfy the cocycle identities
  $$ \aligned
     1 &= g\mstrut _{jk}g_{ik}\inv g\mstrut _{ij}\\
     g_{ij}\inv dg\mstrut _{ij} &= \alpha \mstrut _j-\alpha \mstrut
     _i.\endaligned $$
Notice that the curvature of~$L$ is the 2-form~$d\alpha _i$ for any~$i$.  The
generalization to a 2-circle bundle with connection is straightforward.  The
Deligne cocycle consists of
  $$ \gathered
     g_{ijk}\:U_i\cap U_j\cap U_k\longrightarrow \TT\\
     \alpha _{ij}\in i\Omega ^1(U_i\cap U_j)\\
     \alpha _i\in i\Omega ^2(U_i)\endgathered $$
which satisfy the cocycle identities
  $$ \aligned
     1 &= g\mstrut _{jk\ell }g_{ik\ell }\inv g\mstrut _{ij\ell } g_{ijk}\inv\\
     g_{ijk}\inv dg\mstrut _{ijk} &= -\alpha \mstrut _{jk} + \alpha \mstrut
     _{ik} - \alpha \mstrut _{ij}\\
     d\alpha \mstrut _{ij} &= \alpha \mstrut _j - \alpha \mstrut
     _i.\endaligned $$
The curvature is the 3-form $d\alpha _i$ for any~$i$.

It is much more intuitive to think of the total space of a circle bundle, or
line bundle, rather than the \v Cech data~\thetag{18}.  Unfortunately, we
cannot make sense of the total space of a higher line bundle as a manifold.
One possible alternative advocated by Brylinski~\cite{B} uses the language of
sheaves and categories to describe these objects.  We do not describe that
here.

The integration formally looks very similar to the integration of
differential forms.  We integrate over families of manifolds, not just over a
single manifold.  Thus suppose $\pi \:M\to N$ is a fiber bundle of compact
$d$-manifolds which is {\it relatively oriented\/}.  (This means that the
tangent bundle along the fibers of~$\pi $ is oriented.)  Suppose that $\Lk$
is a $k$-circle bundle with connection over~$M$.  Then if the fibers of~$\pi $
are closed manifolds, the integral $\intLk$ over the fibers is a $(k-d)$-circle
bundle with connection on~$N$.  If the fibers are manifolds with boundary,
then $\intLk$~is a section of the $(k-d+1)$-circle bundle with
connection~$\int_{\partial (M/N)}\Lk$ over~$N$.  This integration satisfies a
gluing law which looks like a parametrized version of~\thetag{8}
or~\thetag{16}.  There are also formulas which compute the curvature of the
integral in terms of the integral of the curvature, the characteristic class
of the integral in terms of the integral of the characteristic class, etc.

It is best to illustrate this integration theory in the case where $L$~is an
ordinary (hermitian) circle bundle with connection over~$M$.  If the typical
fiber of $\pi \:M\to N$ is a circle, then $\int_{M/N}L\:N\to\TT$ computes the
holonomy of the connection around the fibers.  The formula which states that
the curvature of the integral equals the integral of the curvature reduces to
the usual formula for the differential of the holonomy as a function of a
loop in~$M$.  If the fiber of~$\pi $ is an interval, then $\int_{M/N}L$~is
the parallel transport along the fiber.  The gluing law is the usual fact
that parallel transport composes when paths are pasted end-to-end.

Now we can state our reinterpretation of the classical Chern-Simons theory.
The basic data~\thetag{1} determines up to isomorphism a 3-circle bundle with
connection over the classifying space~$BG$.\footnote{We fix its curvature to
be the Chern-Weil 4-form~\thetag{4} of the universal connection.} Once and
for all choose a representative 3-circle bundle for this equivalence class.  If
$\pi \:M\to N$ is a parametrized family of manifolds, then a family of
$G$~connections on~$M$ determines a 3-circle bundle~$\Lt$ with connection
on~$M$.\footnote{This involves a choice of additional data, such as a
connection on $\pi \:M\to N$.  We will more fully investigate this
in~\cite{F3}.} The various entries in Table~1 come from computing
$\int_{M/N}\Lt$, where the dimension in the table is the relative dimension
of~$\pi $.  (If this relative dimension is~4, then we integrate the curvature
of~$\Lt$, which is the Chern-Weil 4-form.)  For example, if $\Theta $~is a
connection on a $G$~bundle $P\to X$ over a closed oriented 3-manifold, then
$\Theta $~determines a 3-circle bundle $\Lt_\Theta \to X$.  The Chern-Simons
action~\thetag{11} is
  $$ e^{\tpi S_X(\Theta )} = \int_{X}\Lt_\Theta . \tag{19}$$
If $\eta $~is a connection on a $G$~bundle $Q\to Y$ over a closed oriented
2-manifold, then $\eta $~determines a 3-circle bundle $\Lt_\eta \to Y$ and the
Chern-Simons circle~\thetag{14} is
  $$ L_\eta =\int_{Y}\Lt_\eta . $$
A parametrized version leads to the circle bundle with
connection~\thetag{17}.  Its curvature, and the moment map for the action of
gauge transformations, can be computed from the general integration theory.
If $\Theta $~is a connection on a 3-manifold~$X$ with boundary, then
\thetag{19}~again gives the action, but now it takes values in the
Chern-Simons circle of~$\partial \Theta $.  The gluing law for the
integration reduces to~\thetag{16}.

We can now fill in Table~1 using this integration theory for higher circle
bundles with connection.  Thus attached to a closed oriented 1-manifold~$S$
is a 2-circle bundle with connection $L^{(2)}_{S}\to\conn S$.  Of course,
$S$~is a finite union of manifolds diffeomorphic to~$\cir$.  Fixing $S=\cir$
for definiteness this 2-circle bundle drops to a 2-circle bundle with
connection $L^{(2)}\to G$ over the space of {\it based\/} equivalence classes
of connections on the circle, which is isomorphic to the group~$G$ via the
holonomy. The integration theory implies that its curvature is~\thetag{3}
and its characteristic class is the transgression of~$\lambda $
in~\thetag{1}.  If $\eta $~is a connection on a compact oriented
2-manifold~$Y$ with boundary, then the integral of the 3-circle bundle gives
an element~$L_\eta $ in the 2-circle~$L^{(2)}_{\partial \eta }$ attached to
the restriction of~$\eta $ to~$\partial Y$.  These objects obey a gluing law
coming from the integration theory.  Also, in parametrized families they fit
together smoothly and the Chern-Simons lagrangian determines a connection on
the resulting 2-circle bundle.

\subhead
$\ZZ$ Structures
\endsubhead

The exponential exact sequence
  $$ 1 @>>> \ZZ @>>> \RR @>{\operatorname{exp}}>> \TT @>>> 1 $$
gives us a way to relate structures for the group~$\TT$ (which we have been
discussing) to structures for the group~$\ZZ$.  The integers already enter
Table~1 in dimension~4, where $\lambda _W$~takes integer values for a closed
oriented 4-manifold~$W$.  Quite generally a number $\tau \in \TT$ determines
a {\it $\ZZ$-torsor\/}
  $$ T_\tau  = \{x\in \RR : e^{\tpi x} = \tau \}; $$
that is, $T_\tau $~is a set with a free transitive $\ZZ$~action.  A
connection~$\Theta $ on a closed oriented 3-manifold determines a
$\ZZ$-torsor
  $$ T\mstrut _{\Theta } = T_{e^{\tpi S_X(\Theta )}}. $$
Also, equation~\thetag{6} implies that if $\Theta $~is a $G$~connection on a
compact oriented 4-manifold~$W$ with boundary, then
  $$ \lambda _W(\Theta )\in T_{\partial \Theta }. $$
We can continue in this way to obtain ``higher $\ZZ$~bundles'' analogous to
the ``higher $\TT$~bundles'' or ``higher circle bundles'' discussed
previously.  These admit unique connections since $\ZZ$~is discrete.  They
determine topological invariants of the underlying principal bundles which
are ``higher'' versions of the characteristic number~\thetag{7}.  We
summarize in Table~3.  Notice that the structure is analogous to that in
Table~1 except that the dimensions are shifted up by one and $\TT$~is
replaced by~$\ZZ$.  The table continues to lower dimensions following the
pattern.

\bigskip
\midinsert
{\setbox0\vbox{\kern6pt\hsize 190pt \centerline{A section~$T_X$ of the
``2-$\ZZ$ bundle''~$r^*T_{\bX}^{(2)}$:\hfil} \abovedisplayskip=6pt
\belowdisplayskip=6pt $$ \CD
r^*T^{(2)}_{\bX}  @>>> T^{(2)}_{\bX}\\ @VVV @VVV \\ \conn X @>r>>
\conn{\bX}\endCD$$}
\setbox1\vbox{\kern6pt\hsize 190pt  \centerline{A section~$\lambda _W$ of the
$\ZZ$ bundle~$r^*T_{\bW}$:\hfil} \abovedisplayskip=6pt
\belowdisplayskip=6pt $$ \CD
r^*T_{\bW}  @>>> T_{\bW}\\ @VVV @VVV \\ \conn W @>r>>
\conn{\bW}\endCD$$}
\offinterlineskip \tabskip = 0pt
  \halign{
	\vrule\enspace\hfil#\hfil\enspace\vrule\hskip2pt
        \vrule&\enspace\hfil#\hfil\enspace
	&\vrule\enspace\hfil#\hfil\enspace\vrule\cr
\noalign{\hrule}
\titlestrut{\bf dim}&{\bf closed manifold}&{\bf compact manifold with
boundary} \cr
\noalign{\hrule}
\vphantom{\vrule height 2pt}&&\cr \noalign{\hrule}
\entry 4:{A function $\lambda_W:\conn W\to\ZZ$}:{$\vcenter{\box1}$}
\noalign{\hrule}
\entry 3:{A $\ZZ$ bundle $T_X\to\conn X$}:{$\vcenter{\box0}$}
\noalign{\hrule}
\entry 2:{$\vcenter{\vbox{\kern6pt\hsize 200pt {A ``2-$\ZZ$
bundle'' or ``bundle of $\ZZ$~gerbes''}
\vskip12pt\centerline{$T^{(2)}_Y\to\conn Y$}\kern6pt}}$}:{$\cdots$}
\vphantom{\vrule height2pt}&&\cr \noalign{\hrule}}
}
\nobreak
\bigskip
\centerline{Table~3: $\ZZ$ structures from the Chern-Simons lagrangian}
\medskip
\endinsert

\subhead
Topological Quantum Field Theory
\endsubhead

Witten introduced topological quantum field theories---a $2+1$~dimensional
theory based on the Chern-Simons action~\cite{W1} and a $3+1$~dimensional
theory which produces Donaldson invariants of 4-manifolds~\cite{W2}---which
give quantum analogs of Table~1 and Table~3.  We summarize them in Table~4
and Table~5.  The basic data for these theories is~\thetag{1} and now the
entries in the tables are supposed to be topological invariants.  These
invariants are motivated by the path integral and the theory of canonical
quantization.  (BRST considerations and supersymmetry enter into the
$3+1$~dimensional theory as well.)  Such considerations are not
mathematically rigorous, and more importantly are not well-understood within
mathematics.  For these particular topological quantum field theories there
is much mathematical work (not involving the path integral) which justifies
many of the predictions made by the path integral.  As we have emphasized
throughout, a characteristic feature of both classical and quantum field
theories which is of great interest in mathematics is {\it locality\/}, as
reflected in gluing laws.  Each row in Table~4 and Table~5 obeys a gluing law
which we now describe briefly.

\bigskip
\midinsert
{\offinterlineskip \tabskip = 0pt
  \halign{
	\vrule\enspace\hfil#\hfil\enspace\vrule\hskip2pt
        \vrule&\enspace\hfil#\hfil\enspace
	&\vrule\enspace\hfil#\hfil\enspace\vrule\cr
\noalign{\hrule}
\titlestrut{\bf dim}&{\bf closed manifold}&{\bf compact manifold with
boundary} \cr
\noalign{\hrule}
\vphantom{\vrule height 2pt}&&\cr \noalign{\hrule}
\entry 3:{$\vcenter{\vbox{\kern6pt\hsize 260pt \centerline{A number
$Z_X\in \CC$}\vskip12pt\centerline{(``partition
function'')}\kern6pt}}$}:{An element $Z_X\in E(\bX)$}
\noalign{\hrule}
\entry 2:{$\vcenter{\vbox{\kern6pt\hsize 260pt \centerline{A finite
dimensional complex inner product space
$E(Y)$}\vskip12pt\centerline{(``quantum Hilbert space'')}\kern6pt}}$}:{An
object $E(Y)\in \Cal{C}_{\partial Y}$}
\noalign{\hrule}
\entry 1:{A ``tensor category'' or ``2-vector space'' $\Cal{C}_S$}:{$\cdots$}
\vphantom{\vrule height2pt}&&\cr \noalign{\hrule}}
}
\nobreak
\bigskip
\centerline{Table~4: Quantum Chern-Simons Theory}
\medskip
\endinsert

\bigskip
\midinsert
{\offinterlineskip \tabskip = 0pt
  \halign{
	\vrule\enspace\hfil#\hfil\enspace\vrule\hskip2pt
        \vrule&\enspace\hfil#\hfil\enspace
	&\vrule\enspace\hfil#\hfil\enspace\vrule\cr
\noalign{\hrule}
\titlestrut{\bf dim}&{\bf closed manifold}&{\bf compact manifold with
boundary} \cr
\noalign{\hrule}
\vphantom{\vrule height 2pt}&&\cr \noalign{\hrule}
\entry 4:{$\vcenter{\vbox{\kern6pt\hsize 260pt \centerline{A number
$Z_X$}\vskip12pt\centerline{(Donaldson invariant)}\kern6pt}}$}:{An element
$Z_W\in E(\bW)$}
\noalign{\hrule}
\entry 3:{$\vcenter{\vbox{\kern6pt\hsize 260pt \centerline{A finite
dimensional module $E(X)$}\vskip12pt\centerline{(Floer
homology)}\kern6pt}}$}:{??}
\noalign{\hrule}
\entry 2:{??}:{$\cdots$}
\vphantom{\vrule height2pt}&&\cr \noalign{\hrule}}
}
\nobreak
\bigskip
\centerline{Table~5: Quantum Donaldson Theory}
\medskip
\endinsert

Let's first consider Table~4.  The partition function~$Z_X$ of a closed
oriented 3-manifold is a topological invariant, as predicted by the path
integral.  This has been proved by Reshetikhin-Turaev~\cite{RT} and others.
There has also been much mathematical activity investigating the vector
spaces~$E(Y)$ attached to 2-manifolds.  The quantum gluing law for
3-manifolds states that
  $$ Z_{X}= \( Z_{X_1}, Z_{X_2} \)\mstrut _{E(Y)} $$
if 3-manifolds $X_1,X_2$ are glued along a common boundary~$Y$ to obtain~$X$.
This is the standard gluing law for the path integral in a quantum field
theory.  The notion that a ``tensor category'' should be attached to a
1-manifold has been advanced by many people, including David Kazhdan and
Graeme Segal.  (Others, such as Ruth Lawrence, advocate the use of ``2-vector
spaces''.)  Such a category has a finite number of equivalence classes,
called {\it labels\/}, and for a 2-manifold~$Y$ we can write
  $$ E(Y) \cong \bigoplus_{\alpha }E(Y,\alpha )\otimes E_\alpha , $$
where $\alpha $~runs over the labels on~$\bY$.  Here $E(Y,\alpha )$~is an
ordinary finite dimensional inner product space and $E_\alpha $~is an object
in~$\Cal{C}_{\partial Y}$.  The gluing law for 2-manifolds can be expressed
in terms of the tensor category or in terms of the labels.  The latter gluing
law takes the form
  $$ E(Y)\cong \bigoplus_{\alpha } w_\alpha \cdot E(Y_1,\alpha ) \otimes
     E(Y_2,\alpha ). \tag{20}$$
This gluing law is one version of the {\it Verlinde formula\/}.  The
numerical factors~$w_\alpha $ which come into~\thetag{20} multiply the inner
product on the Hilbert space, and they make this isomorphism an {\it
isometry\/} of Hilbert spaces.  See~\cite{FQ} for details in the case of a
finite gauge group and ~\cite{Wa} in the case of a continuous gauge group.

The partition function in Table~5 is the Donaldson invariant of a
4-manifold~\cite{D}.  The quantum Hilbert space of a 3-manifold is the Floer
homology~\cite{Fl}, ~\cite{Br}, and the relative Donaldson invariant takes
values in the Floer homology of the boundary.  The base ring here is
presumably some subring of the complex numbers, or even of the rational
numbers.  The associated gluing law has been the driving force behind some
very recent work of Gompf-Mrowka~\cite{GM} and Fintushel-Stern~\cite{FS}
which produces a menagerie of exotic 4-manifolds.  Mathematicians are now
beginning to contemplate the Floer homology of a 3-manifold with boundary and
the associated gluing law.  This work is directly inspired by the ideas
surrounding the quantum Chern-Simons theory.

The passage from Table~1 to Table~4 is formally integration over the space
(of equivalence classes) of connections.  For a 3-manifold this is the
standard path integral.  For a 2-manifold this is usually explained as
canonical quantization, but perhaps one should also think of this as some
sort of integration process.  After all, it has a gluing law characteristic
of all integration processes.  Also, in the integration theory of higher line
bundles we obtain a complex line as the result of integration, so why not
have an integration process which gives a complex Hilbert space?  Indeed, one
could formally think of integrating the Chern-Simons lines over the space of
connections by taking a direct sum of lines.\footnote{Notice that it is the
Chern-Simons {\it lines\/}, not {\it circles\/}, which can be summed.} In the
case of a finite gauge group this direct sum, appropriately weighted by a
measure on the space of connections, is precisely the quantum Hilbert space.
Continuing, one would speculate that in the next dimension there is an
integration process whose result is a 2-vector space.  {\it Et cetera\/}.

The passage from Table~3 to Table~5, if indeed one should view Table~3 as
some sort of classical version of Table~5, also formally looks like
integration.  The classical theory as presented in~\cite{W2} involves
supersymmetry, and then the integration is the (formal) Berezin integral.
Perhaps there are other possibilities.

But these are only speculative remarks.  At the end of the day we are left
with the same mystery that has always been there for mathematicians---the
path integral.  While we have found some novel integrations in finite
dimensions related to classical topological actions, the infinite dimensional
integrals of the quantum theory remain a mystery.  These topological field
theories offer a rich mathematical structure, and they are closely connected
to many parts of mathematics.  Perhaps they will be an important clue towards
understanding the integration processes of quantization in general.

\enddocument